\documentclass{aa}  

\usepackage{graphicx}
\usepackage{txfonts}

\usepackage{color}
\usepackage{multirow}

\begin{document} 

\title{Photometric classification of QSOs from ALHAMBRA survey
       using random forest}
   
\author{Benjam\'in Arroquia-Cuadros
        \and
        N\'estor S\'anchez
        \and
        Vicent G\'omez
        \and
        Pere Blay
        \and
        Vicent Martinez-Badenes
        \and
        Lorena Nieves-Seoane}
        
\titlerunning{Photometric classification of QSOs with random forest}

\authorrunning{Arroquia-Cuadros et al.}

\institute{Universidad Internacional de Valencia (VIU),
           C/Pintor Sorolla 21, E-46002 Valencia, Spain}

\date{Received ...; accepted ...}

\abstract
{Given the current big data era in Astronomy, machine learning based methods have being applied over the last years to identify or classify objects like quasars, galaxies and stars from full sky photometric surveys.}
{Here we systematically evaluate the performance of Random Forests (RF) in classifying quasars using either magnitudes or colours, both from broad and narrow-band filters, as features.}
{
The working data consists of photometry from the ALHAMBRA Gold Catalogue
that we cross-matched with the Sloan Digital Sky Survey (SDSS) and with the Million Quasars Catalogue (Milliquas)
for objects labelled as quasars, galaxies or stars.
A RF classifier is trained and tested to evaluate the effect on final accuracy and precision of varying the free parameters and the effect of using narrow or broad-band magnitudes or colours.}
{
Best performances of the classifier yielded global accuracy and quasar precision around $0.9$. Varying model free parameters (within reasonable ranges of values) has no significant effects on the final classification. Using colours instead of magnitudes as features results in better performances of the classifier, especially using colours from the ALHAMBRA Survey. Colours that contribute the most to the classification are those containing the near-infrared $JHK$ bands.}
{}


\maketitle

\section{Introduction}

In the current big data era, astronomers have to deal with massive amounts of both photometric and spectroscopic data, and the volume of these data will increase even further with current and upcoming surveys. It is generally not possible to identify and/or characterise millions of objects using non-automated techniques and, for this reason, the number of (semi-)automated methods and techniques has been increasing during the last years. In this context, automated classification of sources from wide-field photometric surveys becomes especially relevant because, even though spectroscopic classification should be more accurate and reliable, it is very time-consuming to obtain spectroscopic data. In contrast, multiband photometry can display the overall shape of a spectrum and it is relatively fast to classify objects by using magnitudes and colours criteria.

Over the last years, machine learning based methods have being applied to classify and measure properties of objects using photometric information (magnitudes and/or colours) from wide-field surveys such as the SDSS \citep{York2000}, the Wide-Field Infrared Survey Explorer \citep[WISE;][]{Wright2010} or the Two Micron All Sky Survey \citep[2MASS;][]{Skrutskie2006}. For instance, different algorithms based on decision trees \citep{Suchkov2005,Ball2006,Vasconcellos2011,Li2021,Nakoneczny2021,Cunha2022} or RF \citep{Carrasco2015,Bai2019,Nakoneczny2019,Schindler2019,Clarke2020,Guarneri2021,Nakazono2021,Nakoneczny2021,Wenzl2021},
Support Vector Machines \citep[SVM;][]{Peng2012,Kovacs2015,Krakowski2016,Wang2022}, artificial neural networks \citep[ANN;][]{Yeche2010,Makhija2019,Khramtsov2021,Nakoneczny2021} and k-nearest neighbours \citep[KNN;][]{Khramtsov2021} have been used to classify sources into stars, galaxies and quasars (QSOs). These studies fall into the so-called supervised learning algorithms, in which a subsample of reliable labelled objects is selected and used to train, optimise and test the algorithm's performance. Analyses and comparisons of different methods have been performed by several authors \citep{Bai2019,Nakazono2021,Nakoneczny2021,Wang2022} and some of them seem to indicate that RF tends to show better results in terms of metrics such as accuracy or precision (beyond the computing time required for processing), but this question is still subject to debate. For instance, \citet{Bai2019} found that RF had, on average, higher accuracy than KNN and SVM, contrary to the results of \citet{Wang2022} who found that SVM had better performance than RF. However, these differences may be more related to factors such as the sample (specially the size and quality of the training sample) or the number or type of the used features (magnitudes, colours or different combinations of these and other features) than to the algorithms themselves. In general, we would expect that larger training samples and/or higher number of features imply better results, but this may lead to an overfitting in which the learning patterns work well on the training data only. The search of the optimal strategy may produce dissimilar approaches as using only some broad band magnitudes as features \citep[e.g.][]{Nakazono2021}, or a set of 83 features including magnitudes, colours, ratios of magnitudes and other morphological classifiers \citep{Nakoneczny2021}, or even combining 32 different machine learning models based on three different algorithms \citep{Khramtsov2021}. At this point it is not straightforward to figure out what the best approach is, and this is one of the goals of the present study.

From a photometric point of view, a comparison of the shape of the objects' spectra, necessary to classify them, should be done through their colours and not through their apparent magnitudes. That is why standard classification methods are based on cuts in colour-colour diagrams \citep[see, for instance,][]{Glikman2022}. However, some authors have incorporated partially or exclusively magnitude information as features for the RF classifier \citep[for instance,][]{Clarke2020,Khramtsov2021,Nakazono2021} with apparent good results. This is likely because decision trees in a RF compare features with each other and, in some way, they are working with colours. In fact, \citet{Clarke2020} observed that using differences among magnitudes with a given band instead of magnitudes themselves did not improve the classifier performance, which in principle is a somewhat unexpected result. Another important aspect to consider is the importance of the bandwidth. It seems reasonable to claim that narrow-band information should yield better results than a set of broad-band filters. \citet{Nakazono2021} obtained their best results by combining narrow-band and broad-band magnitudes. The point is that the use of other additional features in the RF, apart from magnitudes/colours, and different techniques or datasets by different authors makes it difficult to draw robust conclusions on these questions. In this work we have centred our analysis on RF because it is one of the most used algorithms due to its computational eﬃciency and simplicity both from the training and from the interpretation points of view, and we have focused on evaluating the real effect of free parameter variations and the importance of using magnitudes or colours as input features as well as the importance of the band widths. Sections~\ref{data} and \ref{clasificador} describe the used data and classifier (RF). In Section~\ref{resultados} we discuss the results and the effects of using different parameters, features or data sets. A direct comparison among different classifiers is presented in Section~\ref{comparison}. Finally, the main conclusions are summarised in Section~\ref{conclusiones}.

\section{Data sets and preprocessing}
\label{data}

Our main working sample comes from the ALHAMBRA Survey \citep{Moles2008} that observed several regions of the sky with the 3.5m telescope at Calar Alto (CAHA, Spain). The distinguishing characteristic of this survey is that it used a system of 20 contiguous, non-overlapping, equal width ($\sim 300$ \AA) filters, covering the optical range ($3500-9700$ \AA), plus the standard broadband $JHKs$ filters. In particular, we use the ALHAMBRA Gold Catalogue \citep{Molino2014}, which contains PSF-corrected photometry for 441303 sources (galaxies, stars and AGN candidates) spread over an effective area of $\sim 3$ deg$^2$ and completed down to a magnitude $I \sim 24.5$ AB. The details about the observations, data reduction and catalogue construction can be found in \citet{Molino2014}.

\begin{figure}
\centering
\includegraphics[width=0.5\textwidth]{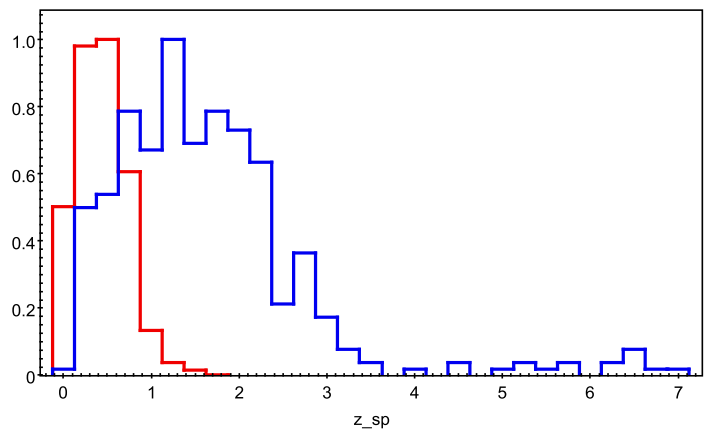}
\caption{Normalised histogram of spectroscopic redshift ($z\_sp$) distribution for the used sample of galaxies (red line) and QSOs (blue line).}
\label{fig:histoz}
\end{figure}

In order to define appropriate training and testing datasets we used SDSS \citep{York2000}. Its Photometric Catalogue DR17 \citep{Sloan22} includes reliable spectroscopic classes for galaxies, stars and QSOs. We cross-matched the ALHAMBRA catalogue with SDSS using a matching radius of 1 arcsec in positions and we retained all the ALHAMBRA and SDSS magnitudes. 
We also cross-matched the last released version of the Milliquas Catalogue \citep[v7.8;][]{Milliquas21} with both ALHAMBRA and SDSS catalogues in order to increase the number of QSOs with magnitudes in these bands.
After removing duplicate entries and sources having no magnitude measurements in any band we obtained a working dataset consisting of
2621 sources.
We are aware that $\sim 10^3$ objects is a relatively small sample size, but as we will see, it is large enough to allow us to reach robust conclusions.
Within this sample, there are 516 stars of different spectral types. Some of these stars are early-type OB (9) or A (39) stars although most of them are of later spectral types F(164), G (36), K (125) and M (124), the remaining being of other classes like carbon stars and white dwarfs. A total of 1347 sources are classified as galaxies, most of them ($\simeq 90$\%) being normal galaxies and the rest being active subclasses such as AGNs or starburst galaxies. Finally, 758 sources are classified as QSOs either by SDSS or by Milliquas. Figure~\ref{fig:histoz} shows the redshift distribution of QSOs from this sample, which it is predominantly in the range $0 \lesssim z \lesssim 3.5$ whereas for galaxies it is in the range $0 \lesssim z \lesssim 1.1$.
The number of correct or missed classifications per class may depend, among other things, on the ranges of redshift values \citep{Clarke2020}. However, our working sample is not large enough to make a detailed analysis of this issue.

\section{Random Forest classifier}
\label{clasificador}

In this work we use a RF classifier \citep{Breiman2001}, which consists in applying a set of independent decision trees to perform the classification. Decision trees are among the most used machine learning algorithms because they are simple and easy to interpret in comparison with other black box models such as artificial neural networks. However, this advantage is counteracted by its lack of robustness: small changes in the training data may lead to big changes in the final results. RFs overcome this limitation because they use many decision trees that are trained on random subsets of the training data, and the final classification is more robust because it is the result of the full set of learning processes and not only one tree.

We use the Scikit-Learn\footnote{https://scikit-learn.org} \citep{Pedregosa2011} library to build our models. The two main parameters when using a RF classifier are the number of trees in the forest ($n\_estimators$) and the size of the random subset of features to be used ($max\_features$). In principle, the larger $n\_estimators$ the better the results, but there is some $n\_estimators$ value beyond which the results do not change significantly. The default value for $n\_estimators$ in the Scikit-Learn library is $n\_estimators=100$, which we will keep as a ``reference'' value (the effect of the free parameters on the results is discussed in Section~\ref{sec_parameters}). According to the Scikit-Learn tutorial, an empirical good value of $max\_features$ for classification tasks is the square root of the total number of available features. For the case of the ALHAMBRA survey (with 23 filters) this corresponds approximately to $max\_features = 5$, that will be our reference value for this parameter. Another important parameter is the maximum allowed depth of the trees ($max\_depth$). By default, Scikit-Learn sets this value to {\it none}, i.e. it allows to fully develop the trees. For the moment we choose $max\_depth=10$ as the reference value and we leave the discussion on the free parameters to Section~\ref{sec_parameters}.

There are other less relevant parameters, such as the minimum number of samples required to split an internal node of the tree, the minimum impurity decrease necessary to split a node, or the function used to measure the quality of the split, either {\it gini} (the default criterion in Scikit-Learn) or {\it entropy}. We ran several tests in which we varied these parameters and, beyond computing performance, the final results are statistically similar so that, in general, we kept these free parameters unchanged to their default (Scikit-Learn) values.

Regarding data samples, we balanced the sizes of the different classes to prevent any bias in the results
and we used a cross-validation splitting strategy (we kept the default 5-fold cross validation given by Scikit-Learn) to train and evaluate the performance of the RF classifier. With these considerations, the classifier had a total of $\sim 1240$ objects to be trained.
Although increasing the training sample should result in a better classification, the optimal training size (i.e. the minimum size leading to a ``good'' classification) depends on several factors including, for instance, the number of input features \citep{Maxwell2018}. However, in comparison with other classifiers, RF seems to show a negligible decrease in its overall accuracy when training samples are reduced to sizes as small as $\sim 300$ samples \citep{Ramezan2021}.
A key aspect of our approach is to keep unaltered the properties of the training and testing samples in order to be able to evaluate the real effect on the classification of using magnitudes instead of colours, or broad instead of narrow bands as features, or different classification algorithms.

\section{Results and discussion}
\label{resultados}

In this section we discuss our main results regarding the effect of the free parameters and the effect of using different features and data sets on the classification performance. We analyse in more detail the results when using as features colours from the ALHAMBRA Survey. In order to evaluate the global performance of the classifier we use the accuracy ($AC$) as main metric, which is a simple and direct measure giving the fraction of well-classified objects (stars, galaxies and QSOs), that is the number of true positives and true negatives divided by the total number of objects.
We also calculate and show other metrics of interest for each object class, such as precision (i.e. the fraction of positive predictions that are actually positive), recall (the fraction of positive data that is predicted to be positive), and F1-score (the harmonic mean of precision and recall). We pay special attention to the
QSO precision ($QP$), i.e. the fraction of QSOs that are well-classified, because we are particularly interested in generating a sample of QSO candidates as clean as possible from stars and galaxies. 
For each classifier execution, we determine mean values 
and uncertainties associated to each metric, which is important for validating whether metric differences are statistically significant or not.

\subsection{Free parameter effects}
\label{sec_parameters}

As mentioned before, our reference cases are the ones for which $criterion = gini$, $n\_estimators = 100$, $max\_features = 5$ and $max\_depth=10$. If the RF classifier is executed using these parameters and the $23$ bands of the ALHAMBRA photometric catalogue, we obtain that $AC = 0.779 \pm 0.007$\footnote{We are expressing associated uncertainties as one standard deviation.} and $QP = 0.743 \pm 0.009$. These values are smaller than those found by other authors. For example, \citet{Clarke2020} obtained $QP \simeq 0.89-0.96$, depending on the used features (SDSS or WISE magnitudes, or a combination of both), for a sample of $\sim 1.5$ million spectroscopically confirmed sources. Differences in performance metrics come, among other things, from the difference in sample sizes. Obviously, final performances would improve if additional features were incorporated, such as for instance the photometric redshift estimations available in the ALHAMBRA Gold Catalogue \citep{Molino2014} but, in any case, the exact range of obtained metrics is not of crucial importance since we are interested in evaluating the {\it relative} contributions of using magnitudes and colours of different band widths.

\begin{figure*}
\centering
\includegraphics[width=0.33\textwidth]{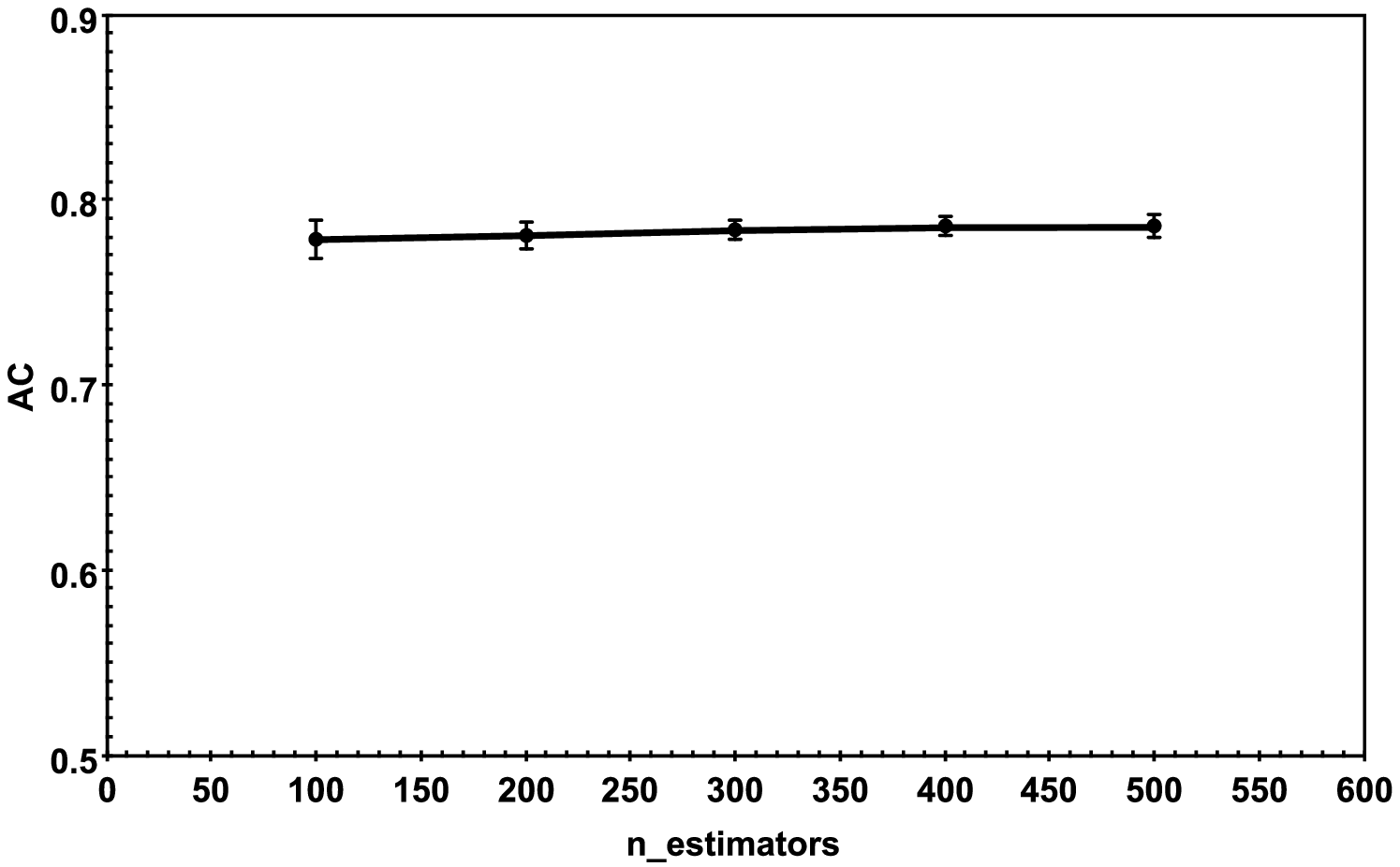}
\includegraphics[width=0.33\textwidth]{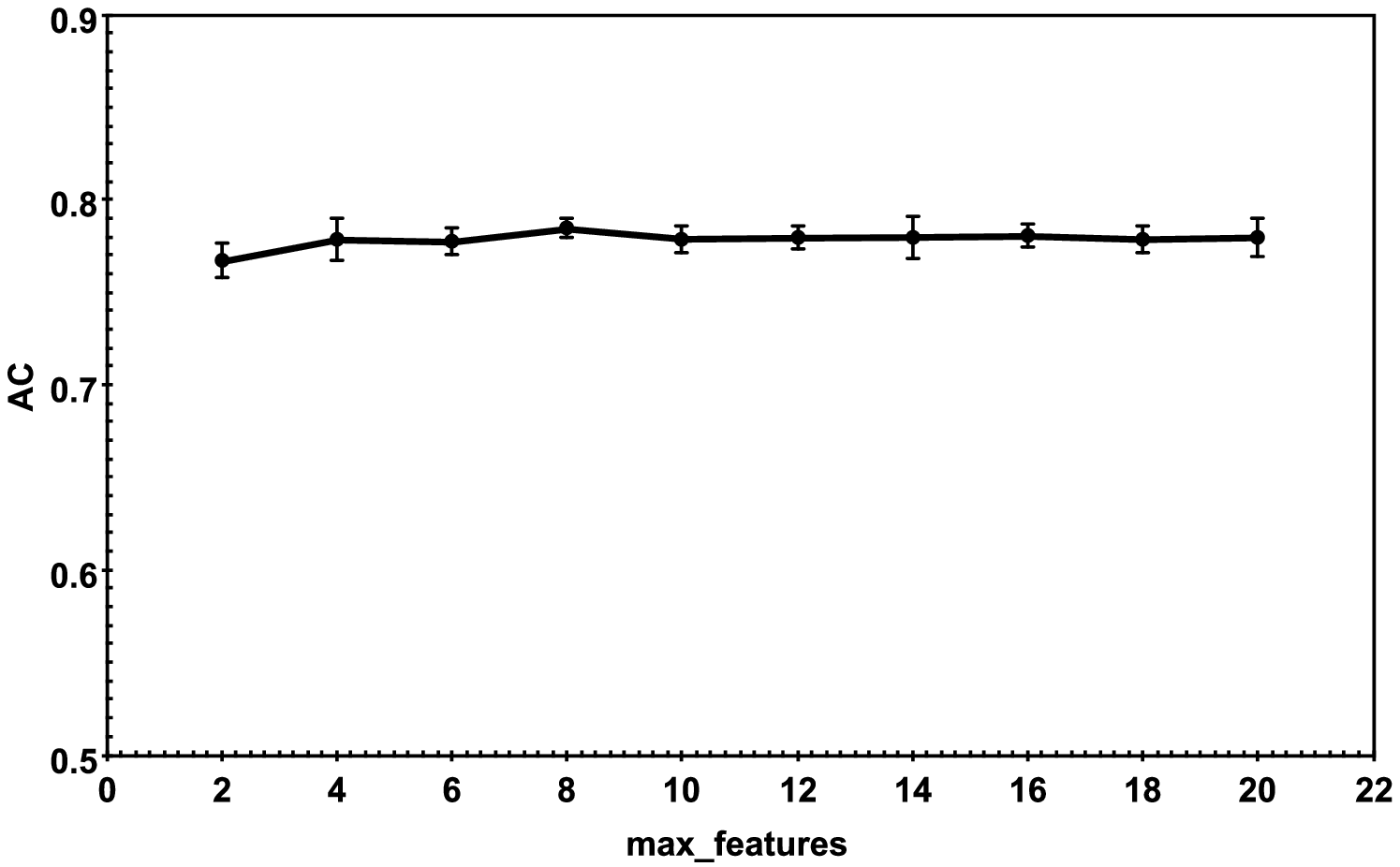}
\includegraphics[width=0.33\textwidth]{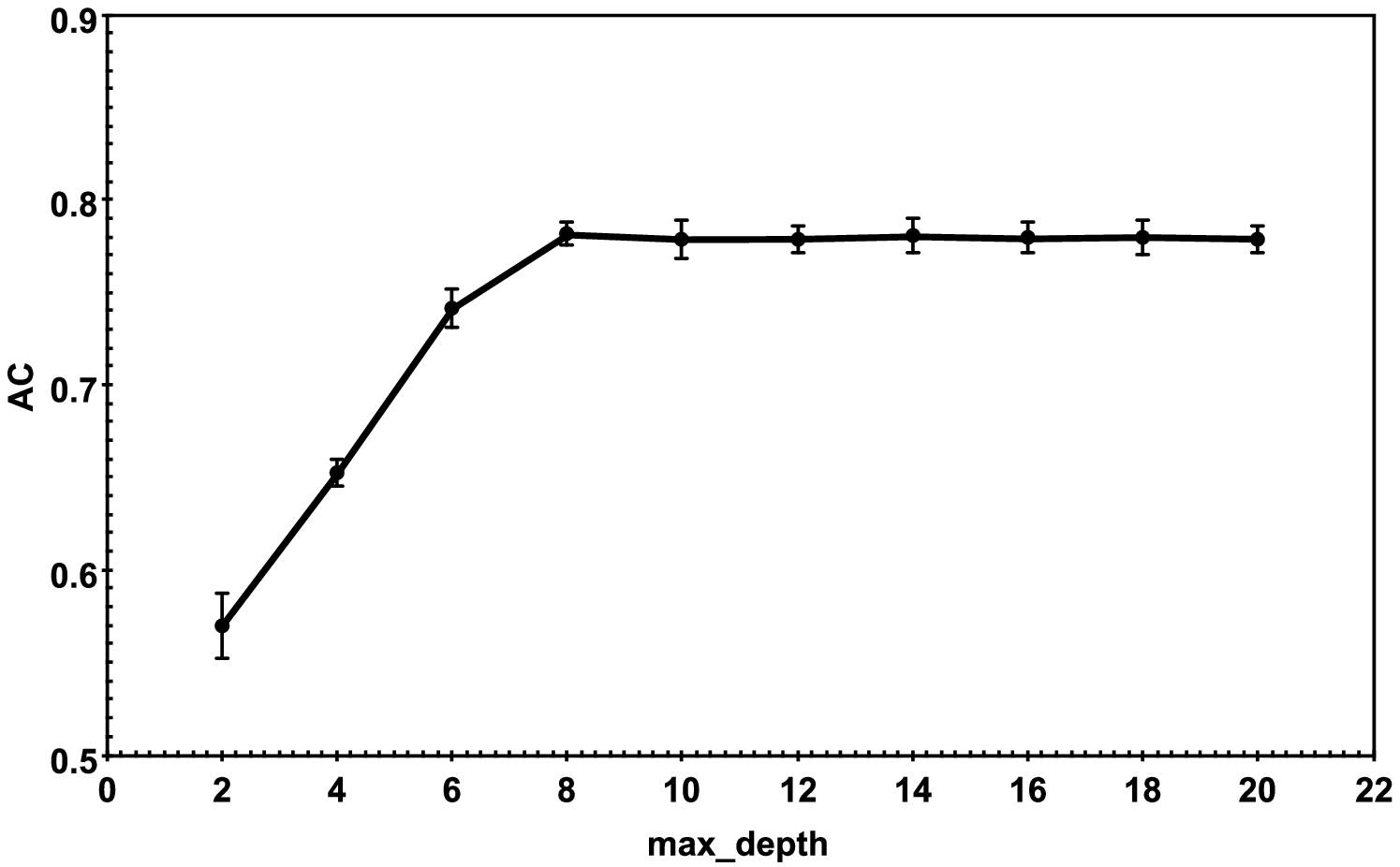}
\caption{Global accuracy ($AC$) as a function of free parameters $n\_estimators$ (left panel), $max\_features$ (central panel) and $max\_depth$ (right panel) when the RF classifier is applied with magnitudes from ALHAMBRA as features. In each case, the rest of the free parameters are fixed to their reference values (see text). Error bars correspond to three times the estimated standard deviations.}
\label{fig:parametros}
\end{figure*}

Figure~\ref{fig:parametros} shows the effects of free parameters variations on the obtained global accuracy.
We see that, even though associated uncertainties decrease, $AC$ remains practically unchanged as the number of trees in the forest increases. The same occurs with the maximum number of features used by the RF, where no significant differences are observed between using almost the full set or only two features (photometric bands). This is not a surprising result because the RF classifier found one relevant feature, the $K$ band magnitude, with an importance index of $0.14$, whereas the rest of features had smaller importance indices in the range $0.02-0.06$. For the case of the maximum trees depth, above a reasonable threshold value ($max\_features \simeq 8$) we obtain that, again, accuracy remains roughly the same.

We wonder what would happen if we use the standard strategy of searching for the optimal combination of free parameters that yields the best classification. For this, we applied the Scikit-Learn tool {\it GridSearchCV} that considers a grid with all possible parameter combinations. 
We spanned the ranges $n\_estimators=100-500$, $max\_depth=2-20$, $max\_features = 2-20$ and $criterion=gini,entropy$, and the optimal set of parameters obtained was $n\_estimators=100$, $max\_depth=10$, $max\_features = 18$, and $criterion=entropy$. With these values, the RF yielded an accuracy that is statistically identical to the reference case ($AC=0.779 \pm 0.005$). In fact, when using other features (colours) or datasets (SDSS) we have found no significant differences between reference and optimal cases (Table~\ref{tab_results} in next section).

\subsection{Different features and datasets}

Our main results 
regarding the global performance of the RF classifier
are summarised in Table~\ref{tab_results}, where a comparison is given among the results obtained for different features (either magnitudes or colours), sets of filters (ALHAMBRA or SDSS\footnote{For a proper comparison, in this work SDSS photometry is always used in combination with the $JHK$ bands that are also included in the ALHAMBRA catalogue.}) and sets of free parameters (reference or optimal).
The corresponding metrics for each class are shown in Table~\ref{tab_results2}.
\begin{table}
\caption{Accuracy ($AC$) and its standard deviation (in brackets) obtained for different combinations of features, databases and free parameter sets.}
\label{tab_results}
\centering
\begin{tabular}{c c c c}
\hline
Features & Database & Parameter set & $AC~(\sigma)$ \\
\hline
Magnitudes & ALHAMBRA & Reference & 0.779~(0.007) \\
Magnitudes & ALHAMBRA & Optimal   & 0.779~(0.005) \\
Magnitudes & SDSS     & Reference & 0.769~(0.006) \\
Magnitudes & SDSS     & Optimal   & 0.772~(0.007) \\
Colours    & ALHAMBRA & Reference & 0.878~(0.004) \\
Colours    & ALHAMBRA & Optimal   & 0.900~(0.002) \\
Colours    & SDSS     & Reference & 0.792~(0.006) \\
Colours    & SDSS     & Optimal   & 0.795~(0.004) \\
\hline
\end{tabular}
\end{table}
\begin{table*}
\caption{Precision, recall and F1 metrics with their standard deviations (in brackets) obtained for different combinations of features, databases and free parameter sets, and for each of the classes in the sample: quasars (QSO), galaxies (GAL) and stars (STA).}
\label{tab_results2}
\centering
\begin{tabular}{c c c c c c c c c c c c}
\hline
Features & Database & Parameter set & Class & Precision$~(\sigma)$ & Recall$~(\sigma)$ & F1$~(\sigma)$ \\
\hline
\multirow{3}{*}{Magnitudes} & \multirow{3}{*}{ALHAMBRA} & \multirow{3}{*}{Reference} &
       QSO & 0.743~(0.009) & 0.712~(0.020) & 0.727~(0.012) \\
& & & GAL & 0.727~(0.016) & 0.743~(0.008) & 0.735~(0.009) \\
& & & STA & 0.864~(0.005) & 0.884~(0.006) & 0.874~(0.004) \\
\hline
\multirow{3}{*}{Magnitudes} & \multirow{3}{*}{ALHAMBRA} & \multirow{3}{*}{Optimal} &
       QSO & 0.752~(0.008) & 0.706~(0.012) & 0.728~(0.009) \\
& & & GAL & 0.730~(0.007) & 0.748~(0.012) & 0.738~(0.007) \\
& & & STA & 0.850~(0.009) & 0.885~(0.007) & 0.867~(0.004) \\
\hline
\multirow{3}{*}{Magnitudes} & \multirow{3}{*}{SDSS} & \multirow{3}{*}{Reference} &
       QSO & 0.749~(0.008) & 0.701~(0.013) & 0.724~(0.010) \\
& & & GAL & 0.695~(0.009) & 0.760~(0.011) & 0.726~(0.008) \\
& & & STA & 0.867~(0.006) & 0.848~(0.007) & 0.858~(0.006) \\
\hline
\multirow{3}{*}{Magnitudes} & \multirow{3}{*}{SDSS} & \multirow{3}{*}{Optimal} &
       QSO & 0.757~(0.014) & 0.702~(0.015) & 0.728~(0.011) \\
& & & GAL & 0.699~(0.011) & 0.758~(0.009) & 0.727~(0.009) \\
& & & STA & 0.864~(0.007) & 0.859~(0.011) & 0.862~(0.007) \\
\hline
\multirow{3}{*}{Colours} & \multirow{3}{*}{ALHAMBRA} & \multirow{3}{*}{Reference} &
       QSO & 0.860~(0.007) & 0.839~(0.009) & 0.849~(0.005) \\
& & & GAL & 0.847~(0.007) & 0.835~(0.009) & 0.841~(0.005) \\
& & & STA & 0.923~(0.005) & 0.959~(0.004) & 0.941~(0.003) \\
\hline
\multirow{3}{*}{Colours} & \multirow{3}{*}{ALHAMBRA} & \multirow{3}{*}{Optimal} &
       QSO & 0.885~(0.006) & 0.881~(0.004) & 0.883~(0.002) \\
& & & GAL & 0.885~(0.003) & 0.853~(0.007) & 0.869~(0.004) \\
& & & STA & 0.928~(0.004) & 0.964~(0.005) & 0.945~(0.003) \\
\hline
\multirow{3}{*}{Colours} & \multirow{3}{*}{SDSS} & \multirow{3}{*}{Reference} &
       QSO & 0.750~(0.012) & 0.683~(0.012) & 0.715~(0.009) \\
& & & GAL & 0.719~(0.008) & 0.775~(0.014) & 0.746~(0.008) \\
& & & STA & 0.904~(0.003) & 0.922~(0.006) & 0.913~(0.004) \\
\hline
\multirow{3}{*}{Colours} & \multirow{3}{*}{SDSS} & \multirow{3}{*}{Optimal} &
       QSO & 0.763~(0.008) & 0.675~(0.012) & 0.716~(0.008) \\
& & & GAL & 0.714~(0.008) & 0.773~(0.006) & 0.742~(0.004) \\
& & & STA & 0.904~(0.003) & 0.942~(0.005) & 0.922~(0.003) \\
\hline
\end{tabular}
\end{table*}

There are no significant differences between using the optimal set of parameters estimated by {\it GridSearchCV} and any other reasonable combination of free parameters, including our reference combination that is shown in Table~\ref{tab_results}. Also, if magnitudes from a system with narrower filters as ALHAMBRA are used, the results are marginally better ($AC \sim 0.78$) than using magnitudes from SDSS ($AC \sim 0.77$). However, in spite of the uncertainties, it seems clear that using colours instead of magnitudes results in better performances of the classifier. The best performance ($AC \sim 0.88-0.90$) occurs when colours from a narrow band system as ALHAMBRA are used as features. In this case, the global accuracy is $> 10$ standard deviations higher than using magnitudes as features.

The latter result is, in principle, expected because from a photometric point of view the difference between QSOs and other objects like galaxies and stars should be reflected on their spectral energy distributions (SEDs), and the shape of the SEDs is described basically by colours. The classification algorithm works comparing features and it could be considered that comparing magnitudes is equivalent, in some way, to work with colours. However, our results indicate that a better classification is obtained when colours are used directly as input for the RF classifier. Additionally, colours calculated using narrow bandwidths more accurately describe the SED shape, and that is likely the reason why we get better performances with ALHAMBRA than with SDSS.

Regarding the performance for each class (Table~\ref{tab_results2}), we can see that all the metrics yield significantly higher values and lower associated uncertainties when colours from ALHAMBRA are used as features, which seems to indicate that using narrow band colours as inputs tends to produce more robust results. Moreover, in concordance with results in Table~\ref{tab_results}, there is no or little difference between using the optimal or the reference set of free parameters. For the best results, those corresponding to colours from ALHAMBRA, values of precision, recall and F1-score for the different classes (QSO, GAL, STA) are always in the range $\sim 0.85-0.95$. QSO and GAL results are quite similar whereas STA metrics tend to give better values. In fact, stars always yield better metrics for any combination of features or databases and this is a reasonable result because, in general, star colours in our working sample tend to be different from those of galaxies and QSOs (see, for instance, Figure~\ref{fig:colorcolor}).

\subsection{Results using colours from the ALHAMBRA catalogue}

\begin{figure}
\centering
\includegraphics[width=0.5\textwidth]{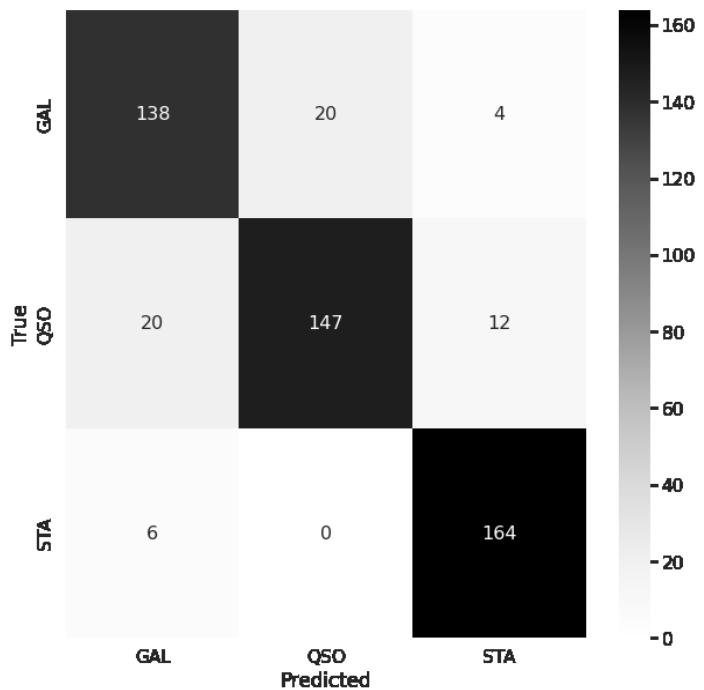}
\caption{Resulting confusion matrix of predicted and true classes for the first random execution of the RF using ALHAMBRA colours as features and the standard set of free parameters.}
\label{fig:confusion}
\end{figure}

\begin{figure}
\centering
\includegraphics[width=0.5\textwidth]{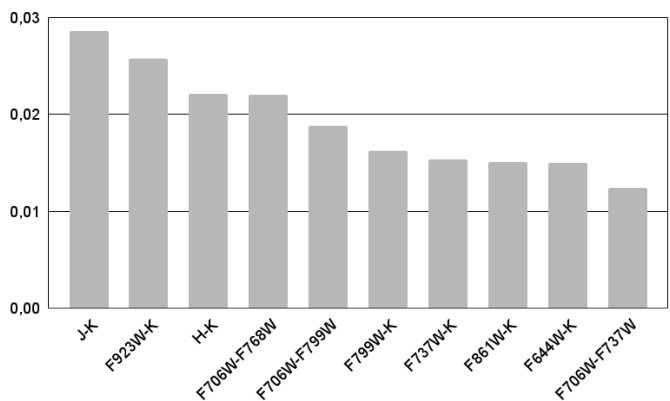}
\caption{Relative importances of the features (colours) for the same execution shown in Figure~\ref{fig:confusion}. For clarity, only the ten most important features are presented.}
\label{fig:colores}
\end{figure}

\begin{figure}
\centering
\includegraphics[width=0.5\textwidth]{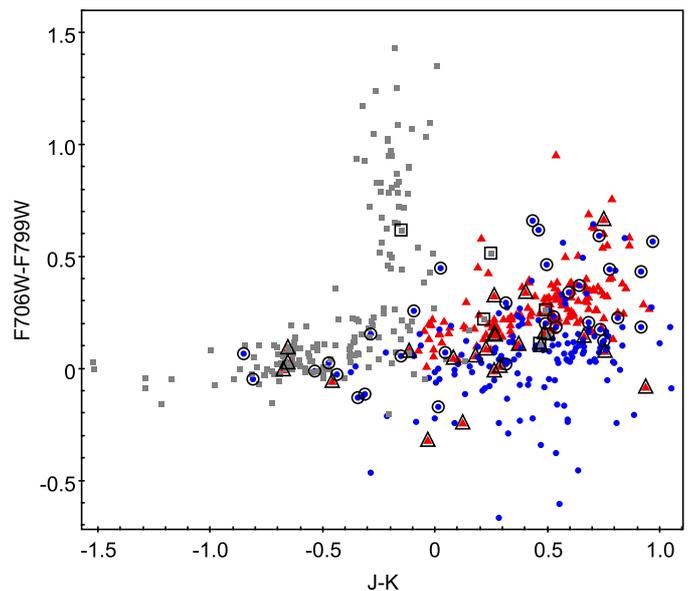}
\caption{Colour-colour diagram, $F706W-F799W$ versus $J-K$, for all stars (grey solid squares), galaxies (red solid triangles) and QSOs (blue solid circles) in the sample, for the same execution shown in Figure~\ref{fig:confusion}. Misclassified sources are also plotted as open symbols for stars (squares), galaxies (triangles) and QSOs (circles).}
\label{fig:colorcolor}
\end{figure}

Our most accurate results were obtained when colours from the ALHAMBRA catalogue were used as features in the RF classifier. Given that there is a total of $23$ filters, the number of available colours is $253$. Mean global accuracy and QSO precision for the reference case were $AC = 0.878 \pm 0.004$ and $QP = 0.860 \pm 0.007$, respectively (Tables~\ref{tab_results} and \ref{tab_results2}). Figure~\ref{fig:confusion} shows the confusion matrix for the first random execution using the reference parameters configuration.
We see that, even for our relatively small training sample size, the model does a good classification. Precisions obtained in this execution for QSO, galaxies and stars were $0.88$, $0.84$ and $0.91$, respectively. As mentioned before, stars were always better classified than QSOs and galaxies. The corresponding feature importances are shown in Figure~\ref{fig:colores}.
Colours that contribute the most to the overall classification are the reddest ones and those using the $K$-band, in particular $J-K$, $F923W-K$, $H-K$, $F706W-F768W$ and $F706W-F799W$ are the five most relevant features. Figure~\ref{fig:colorcolor} shows as an example the $F706W-F799W$ versus $J-K$ plot for our working sample.
We can see that $J-K$ is an useful feature to separate stars because, in general, galaxies and QSOs tend to be redder than stars. This result agrees with that by \citet{Bai2019} who found that IR colours like those involving $JHK$ bands play an important role in Star-Galaxy-QSO classification with RFs. Similarly, \citet{Nakoneczny2021} noted the importance of NIR $JHK$-bands using XGBoost models based on colours and magnitude ratios. Our results also point out the convenience of using IR colours, less affected by dust than optical ones, for classifying QSOs when using only photometric information, and the same conclusion was obtained when we used SDSS and $JHK$ bands exclusively. Misclassified sources (marked as open symbols in Figure~\ref{fig:colorcolor}) tend to be spread over the colour-colour regions where different classes coexist without any obvious bias. 
Misclassified QSOs (open circles in the example shown in Figure~\ref{fig:colorcolor}) represent $\simeq 18$\% of the total QSO sample. From this, those classified as galaxies ($\sim 11$\%) are QSOs that in the colour-colour diagram tend to be located near the region where most of galaxies are found. A similar behaviour occurs for QSOs that are incorrectly classified as stars ($\sim 7$\%). We are particularly interested in knowing what galaxies and stars are misclassified as QSOs because these are the sources that may contribute to contaminate any potential catalogue of QSO candidates. All misclassified stars ($\simeq 3.5$\% of the star sample, plotted as open squares in the example shown in Figure~\ref{fig:colorcolor}) are of spectral type F or later, but these stars are incorrectly identified as galaxies and not as QSOs. Instead, $\simeq 12$\% of the galaxies are wrongly labelled as QSOs, whether they are normal galaxies ($\simeq 7$\%) or starburst galaxies ($\simeq 5$\%) according to SDSS.

\subsection{Results using colours from the SDSS catalogue}

\begin{figure}
\centering
\includegraphics[width=0.5\textwidth]{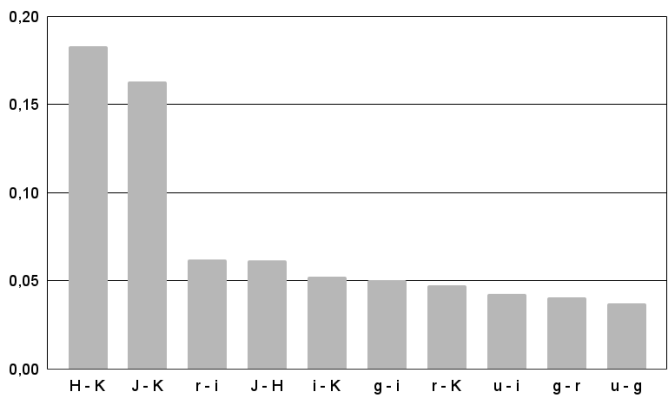}
\caption{Relative importances of the SDSS colours for the reference case. Only the ten most relevant colours are shown.}
\label{fig:coloresSDSS}
\end{figure}

When photometry from SDSS combined with the $JHK$ bands is used by the classifier, metric values tend to be smaller than those for ALHAMBRA photometry. As before, using colours instead of magnitudes yields better performances although not as good as using colours from ALHAMBRA (see Tables~\ref{tab_results} and \ref{tab_results2}). Global accuracy when using SDSS colours is around $AC \simeq 0.79$ and the QSO precision gives $QP \simeq 0.75-0.76$. Stars are once again the best classified class with precision values around $0.9$. As in the previous section, misclassified stars are exclusively classified as galaxies, not as QSOs, whereas $\lesssim 15$\% of the galaxies present in the sample are incorrectly identified as QSOs. These results are very similar to the ones obtained with ALHAMBRA colours. The reason for this is that the main colours contributing to the classification include the $JHK$ bands that are common in both of our databases. The importances of the SDSS colours used as predictors by the classifier is shown in Figure~\ref{fig:coloresSDSS}, where we can see that $H-K$ and $J-K$ clearly stand out as the two most important features, which are also identified as important features from the ALHAMBRA database. The significantly better performance with colours from ALHAMBRA (in comparison with SDSS) is likely related to the additional narrow band colours that more accurately describe the shapes of the different SEDs.

\section{Comparison with different algorithms}
\label{comparison}

Our previous results suggest that a key ingredient for a relatively good photometric classification of QSOs is, apart from having a high-quality training set, to use colours instead of magnitudes as input data. The set of used filters does not necessarily have to be very large but it should include IR bands. Instead, the classification model itself as well as the exact values of its free parameters do not seem to play an important role. In order to verify this, we have run several experiments using other common supervised classifiers. Although the analysis in this section is not as detailed as for RF, we tested different configurations, free parameter sets and parameter optimisation strategies. For a consistent comparison with our previous results for RF, we maintained unaltered the characteristics of the training and testing sets. As input features we used either the full set of available colours from ALHAMBRA or SDSS, or the corresponding ten most relevant ones shown in Figures~\ref{fig:colores} or \ref{fig:coloresSDSS}. The classification algorithms that we studied were:
\begin{itemize}
    \item K-Nearest Neighbours (KNN).
    \item Gradient Boosting (GBoost).
    \item Support Vector Classifier (SVC).
    \item Feedforward Neural Networks (FNN).
\end{itemize}
All the models (except FNN) were implemented with the same previously used Scikit-Learn library \citep{Pedregosa2011} and, in each case, we used k-fold cross-validation with shuffling and calculated 
the global accuracy and the other relevant metrics (precision, recall and F1-score) for each of the classes.
For the case of neural networks we used a single layer FNN implemented with the nnet package on R \citep{Ripley1996, Venables2002} with its default options (logistic output units, least squares fitting) as method for supervised training. The inputs to the neural network were the ten most relevant colours (Figures~\ref{fig:colores} or \ref{fig:coloresSDSS}) and the network was configured with ten input neurons, five neurons in the hidden layer and one output neuron. From the full set of executions with different parameters and criteria specific to each model, at the end we keep and show the best result of each classifier, which is summarised in Table~\ref{tab_results3}.
\begin{table*}
\caption{Global accuracy ($AC$), precision, recall and F1 metrics with their standard deviations $\sigma$ (in brackets) obtained with different algorithms using either ALHAMBRA or SDSS colours.}
\label{tab_results3}
\centering
\begin{tabular}{c c c c c c c}
\hline
Database & Classifier & $AC~(\sigma)$ & Class & Precision$~(\sigma)$ & Recall$~(\sigma)$ & F1$~(\sigma)$ \\
\hline
\multirow{3}{*}{ALHAMBRA} & \multirow{3}{*}{KNN} & \multirow{3}{*}{0.840~(0.025)} &
      QSO & 0.869~(0.073) & 0.708~(0.091) & 0.774~(0.043) \\
& & & GAL & 0.773~(0.065) & 0.858~(0.060) & 0.810~(0.035) \\
& & & STA & 0.893~(0.061) & 0.962~(0.025) & 0.925~(0.031) \\
\hline
\multirow{3}{*}{ALHAMBRA} & \multirow{3}{*}{GBoost} & \multirow{3}{*}{0.864~(0.020)} & 
      QSO & 0.845~(0.043) & 0.801~(0.039) & 0.821~(0.025) \\
& & & GAL & 0.820~(0.058) & 0.839~(0.044) & 0.828~(0.030) \\
& & & STA & 0.923~(0.053) & 0.954~(0.023) & 0.937~(0.027) \\
\hline
\multirow{3}{*}{ALHAMBRA} & \multirow{3}{*}{SVC} & \multirow{3}{*}{0.828~(0.018)} & 
      QSO & 0.797~(0.061) & 0.733~(0.056) & 0.761~(0.031) \\
& & & GAL & 0.789~(0.068) & 0.801~(0.049) & 0.792~(0.041) \\
& & & STA & 0.892~(0.054) & 0.954~(0.023) & 0.921~(0.025) \\
\hline
\multirow{3}{*}{ALHAMBRA} & \multirow{3}{*}{FNN} & \multirow{3}{*}{0.838~(0.014)} & 
      QSO & 0.690~(0.040) & 0.440~(0.050) & 0.540~(0.030) \\
& & & GAL & 0.860~(0.020) & 0.914~(0.008) & 0.884~(0.011) \\
& & & STA & 0.859~(0.018) & 0.930~(0.030) & 0.892~(0.013) \\
\hline
\multirow{3}{*}{SDSS} & \multirow{3}{*}{KNN} & \multirow{3}{*}{0.784~(0.020)} & 
      QSO & 0.709~(0.054) & 0.701~(0.038) & 0.704~(0.037) \\
& & & GAL & 0.722~(0.046) & 0.709~(0.039) & 0.714~(0.029) \\
& & & STA & 0.912~(0.046) & 0.945~(0.032) & 0.927~(0.026) \\
\hline
\multirow{3}{*}{SDSS} & \multirow{3}{*}{GBoost} & \multirow{3}{*}{0.795~(0.017)} & 
      QSO & 0.725~(0.049) & 0.704~(0.046) & 0.713~(0.035) \\
& & & GAL & 0.732~(0.062) & 0.737~(0.055) & 0.734~(0.053) \\
& & & STA & 0.916~(0.032) & 0.943~(0.032) & 0.929~(0.021) \\
\hline
\multirow{3}{*}{SDSS} & \multirow{3}{*}{SVC} & \multirow{3}{*}{0.789~(0.026)} & 
      QSO & 0.738~(0.075) & 0.662~(0.060) & 0.696~(0.055) \\
& & & GAL & 0.714~(0.070) & 0.747~(0.059) & 0.729~(0.058) \\
& & & STA & 0.901~(0.039) & 0.959~(0.020) & 0.928~(0.017) \\
\hline
\multirow{3}{*}{SDSS} & \multirow{3}{*}{FNN} & \multirow{3}{*}{0.871~(0.015)} & 
      QSO & 0.730~(0.040) & 0.650~(0.050) & 0.680~(0.030) \\
& & & GAL & 0.901~(0.017) & 0.915~(0.013) & 0.908~(0.012) \\
& & & STA & 0.880~(0.017) & 0.916~(0.016) & 0.898~(0.015) \\
\hline
\end{tabular}
\end{table*}

We obtained that classifier performances were close to each other and global accuracies are compatible within associated uncertainties. Stars tend to be classified better (higher precisions) than galaxies and QSOs by all the algorithms except FNN. This is likely due to the fact that, in general, intrinsic shapes of star SEDs and therefore their colours differ more than those of galaxies and QSOs (see, for instance, Figure~\ref{fig:colorcolor}).
Global performances for each algorithm are better for ALHAMBRA colours than for SDSS colours. Best overall performance for ALHAMBRA colours is achieved by GBoost with $AC = 0.86$ which is not too far from the obtained RF accuracy ($AC = 0.88-0.90$), not a surprising result considering that RF and GBoost are both based on multiple decision trees although they are trained in different ways. From the point of view of QSO classification, RF with ALHAMBRA colours and the optimal parameter set seems to have a better performance ($QP \simeq 0.89$) than the other methods although, strictly speaking, the difference with KNN, GBoost and SVC is $\lesssim 3$ standard deviations. The exception to these results and behaviours is FNN with a higher accuracy for SDSS colours than for ALHAMBRA colours and a relatively small precision value when using ALHAMBRA colours ($QP \simeq 0.69$),
which is likely associated to the sensitivity of neural networks to small training datasets. FNN results could be improved by optimising the network through its topology, for example with another layer of hidden neurons, but this is beyond the goals of the present work.

\section{Conclusions}
\label{conclusiones}

In this work we have {\it systematically} evaluated the performance of RFs in classifying QSOs, galaxies and stars using either magnitudes or colours, both from broad and narrow-band filters, as features. Our main results are summarised in 
Tables~\ref{tab_results} and \ref{tab_results2}.
The effect of varying free parameters (within reasonable ranges of values), such as the number of random trees, their maximum depths, or the maximum number of used features, is negligible, and the results (global accuracy and class precision) are always the same within their associated uncertainties. Using colours instead of magnitudes as features results in better performances of the classifier, especially using colours from ALHAMBRA. However, colours that contribute the most to the classification are those that contain the $JHK$ bands, in agreement with results found by other authors. The best performance (that using colours from ALHAMBRA) yielded
an accuracy around $0.88-0.90$ and QSO precision of $0.86-0.89$, depending on the used free parameters.
From this work we can conclude that a key point to accurately identifying QSOs from photometric data exclusively is to use a set of colours that includes IR bands and, of course, a good input dataset, whereas the classification model and the exact value of its free parameters are not determinant for obtaining accurate results.
Upcoming large-scale surveys like J-PAS \citep{Benitez2014}, which will observe through $54$ narrow-band filters, seem appropriate to be used as input data for machine learning algorithms that use photometric colours to search for new QSOs. However, given the results of this work, that could be an unnecessary task and it may be enough to use a much smaller subset of colours that includes NIR bands.

\begin{acknowledgements}
We are very grateful to the referee for his/her careful reading of the manuscript and helpful comments and suggestions, which improved this paper.
NS acknowledges financial support from the Spanish Ministerio de Ciencia, Innovación y Universidades through grant PGC2018-095049-B-C21. During this work we have made extensive use of TOPCAT \citep{Taylor2005}.
\end{acknowledgements}

\bibliographystyle{aa}
\bibliography{QSObibliografia}

\end{document}